# New model of the pinning potential barrier in HTc superconductors

J. Sosnowski

Electrotechnical Institute, Pożaryskiego 28, Warsaw, Poland

sosnowski.jacek@wp.pl

Abstract

New model of the pinning potential barrier in multi-layered HTc superconductors has been presented, basing on geometrical approach to the capturing interaction of pancake type vortices with nano-scale defects. Using this model the transport current flow phenomena in these materials, especially current-voltage characteristics and critical current density, have been considered. Details of theoretical analysis are given, including derivation of basic mathematical equations describing the potential barrier in the function of transport current density and initial position of captured pancake vortex. Computer simulation has been performed then of influence of transport current amplitude on potential barrier height for various sizes of pinning centers and initial pancake vortex position as well as influence of fast neutrons irradiation on critical current of HTc layered superconductor.

1. Model presentation

High temperature oxide superconductors are very attractive materials from applied, technical point of view, especially promising in electrical engineering applications. Firstly it is expected and already realized applications of HTc materials in superconducting cryocables, fault current limiters and SMES-s (Superconducting Magnets Energy Storages) as well as superconducting motors more economic, of small dimensions and noiseless. For more efficient use of these unique materials in devices it is necessary however to be more familiar with especial electromagnetic properties of them, caused by multilayered structure leading to characteristic shape of magnetic vortices called now pancake type. Interaction of pancake vortices with structural defects, created for instance by fast neutrons irradiation, considered in the present paper, plays essential function in the transport current flow process through HTc multilayered superconductors and is observed in critical current and current-voltage characteristics. Pancake vortices are most relevant in Bi-2212 superconductors [1-5], where anisotropy effects occur. In the paper it will be analyzed individual interaction of pancake type vortices appearing in the multilayered copper-oxide based HTc superconductors with nano-sized defects, created by fast neutrons irradiation, for geometry of the magnetic field perpendicular to the superconducting layers, through which current flows. Then pancake vortices bring regular form of the short rigid cylinder, while anisotropy effects and Josephson's coupling in present, first approach are neglected. From the thin pancake structure of parallel vortices the flux cutting effects also do not appear, while multi-vortices interaction is in this model taken into account only through analysis of the change of elasticity energy of the vortex lattice during the capturing process. In the model is considered geometry of the ordered nano-defects, which form square lattice repeated in each superconducting layer. More advances geometries and physical cases will be considered in next papers, while present one is devoted to description of the formation of pinning potential energy barrier for individual interaction of pancake vortex with capturing centers of various size and initial states.

An example of the experimental current-voltage characteristics for $Bi_{1.6}Pb_{0.3}Sr_2Ca_2Cu_{3.06}O_8$ high temperature superconducting ceramic with $T_c=112$ K, in static magnetic field and liquid nitrogen temperature is shown in Fig. 1. Sample was prepared by the solid state reaction in furnace in air atmosphere from the oxides of elements and then long time heat treated at $840^0$ C. Phase diagram for this compound was investigated in [6].

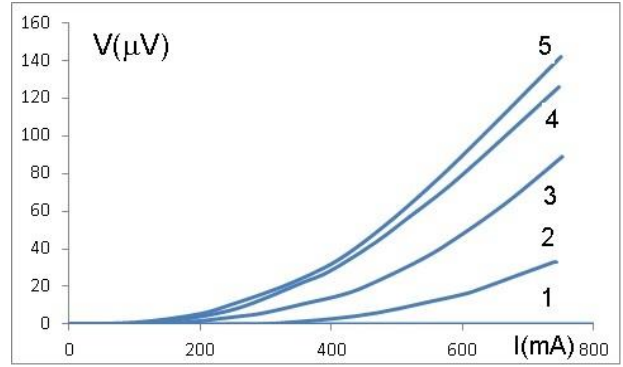

Fig. 1. Experimental current-voltage characteristics for $Bi_{1.6}Pb_{0.3}Sr_2Ca_2Cu_{3.06}O_8$ superconductor in liquid nitrogen temperature for various magnetic fields: (1) B=0, (2) 13,5 mT, (3) 24 mT, (4) 33 mT, (5) 35 mT.

2. Analysis of the pinning potential barrier height $\Delta U$

In real superconductor it should be considered pinning interaction for system of many captured vortices, which total free energy $F$ can be described in the general symbolic form:

$$F(r_1, r_2, r_3, \ldots r_N) = \sum_{i=1}^{N} U(r_i) + \frac{1}{2}\sum_{i \neq j}^{N} F_{inter}(r_i - r_j) - J\emptyset_0 \sum_{i=1}^{N} l_i\, \delta r_i - \sum_{i=1}^{N} \frac{C\delta r_i^2}{2} V_i \quad (1)$$

$U$ is individual pinning potential, while $F_{inter}$ energy of electromagnetic interaction between vortices at position $r_i$ and $r_j$. Summation concerns all vortices transporting flux $\emptyset_0$. Third term is connected with Lorentz force acting on the length $l_i$ and shifting vortex initially being in the position $r_i$ on the distance $\delta r_i$ in the flux creep process, $J$ is current density. Last term describes energy connected with the elasticity forces of the pancake vortices lattice. $V_i$ is volume of deformed lattice during capturing of i-th vortex, while $C$ spring constant of vortex lattice. Equilibrium condition denotes vanishing of the derivative of free energy $F$

$$\frac{\partial F(r_1, \ldots, r_N)}{\partial r_i} = 0 \quad (2)$$

for detailed form of the pinning potential, which exact shape is considered just in this paper. Assumption that pinning center is captured by single pancake vortex only, as is the most frequent case in multilayered HTc superconductors, allows to neglect exchange interaction between vortices and to separate this many-body problem into individual interaction of pancake vortex with capturing center, which analytical solution will be found in the present paper. Others pinning interaction models are given for instance in [7-8].

Analysis of the capturing effects is performed basing on the geometrical approach, which corresponds to Ginzburg – Landau theory in the approximation of lowest order. It has been considered the variation of the energy of system with captured pancake vortex created in perpendicular to superconducting layers magnetic field in the function of vortex declination from initial position under influence of current flow inside of layers, versus defects dimensions. Three initial states configurations have been analyzed. Firstly was considered geometry in which pancake vortex initially is captured on the depth equal to the coherence length, as it shows Fig. 2.

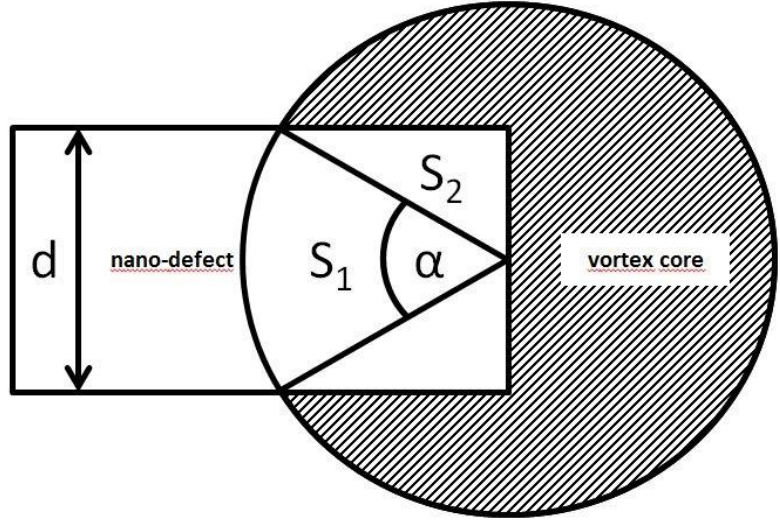

Fig. 2. Scheme of the initial configuration of pancake vortex captured on the nano-sized defect. The symbols meanings is given in the text.

Then the vortex structure is still maintained, including the shielding current distribution. Such approach is useful especially for defect of larger size. Others initial configurations depending on defects properties will be considered in further part. Energy of initial state for captured vortex is given then by the following equation, in which energy scaling has been applied assuming that outside of the capturing center pinning energy vanishes:

$$U(0) = \frac{-\mu_0 H_c^2}{2} l \xi^2 \left( arcsin \frac{d}{2\xi} + \frac{d}{2\xi} \sqrt{1 - (\frac{d}{2\xi})^2} \right) \qquad (3)$$

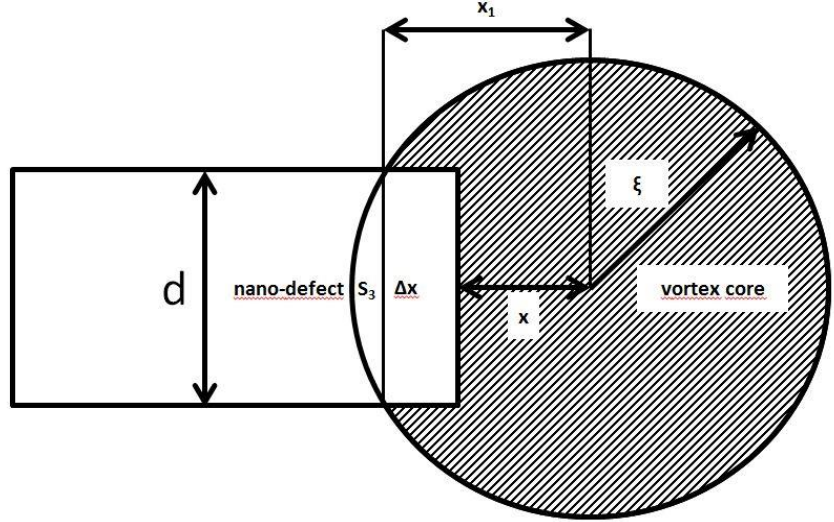

Fig. 3. Geometry of the capturing process for pancake vortex shifted on the length $x$ in the respect to initial state.

where $H_c$ is thermodynamic critical field, $l$ thickness of superconducting layer, $d$ width of pinning center, $\xi$ is coherence length. In derivation of Eq. 3 have been used following quantities received basing on Fig. 2 and given below relations:

$$S_1 = \xi^2 \arcsin\frac{d}{2\xi} \qquad S_2 = \frac{d}{2}\sqrt{\xi^2 - \frac{d^2}{4}} \qquad \alpha = 2\arcsin\frac{d}{2\xi} \qquad (4)$$

For the movement of the pancake vortex center on the length $x$, calculated as the distance between the edge of the nano-sized defect and center of the vortex core, under action of the Lorentz force during the current flow, the following part $S_0$ of the vortex core lies outside of the capturing nano-defect, leading to an enhancement of the normal state energy, as it shows Fig. 3 and describe relations 5-6:

$$S_0 = \pi\xi^2 - \Delta x \cdot d - S_3 \qquad (5)$$

where

$$\Delta x = \sqrt{\xi^2 - \frac{d^2}{4}} - x, \ S_3 = \frac{\alpha\xi^2}{2} - \frac{d\xi}{2}\sqrt{1 - \left(\frac{d}{2\xi}\right)^2}, \ x_1 = x + \Delta x \qquad (6)$$

It means that potential energy for such energy based approach and shift of the pancake vortex on the distance

$$0 < x < x_c = \xi\sqrt{1 - \left(\frac{d}{2\xi}\right)^2} \qquad (7)$$

is given by the relation:

$$U_1(x) = \frac{\mu_0 H_c^2}{2} l \left(dx - \xi^2 \arcsin\frac{d}{2\xi} - \frac{d\xi}{2}\sqrt{1 - \left(\frac{d}{2\xi}\right)^2}\right) \qquad (8)$$

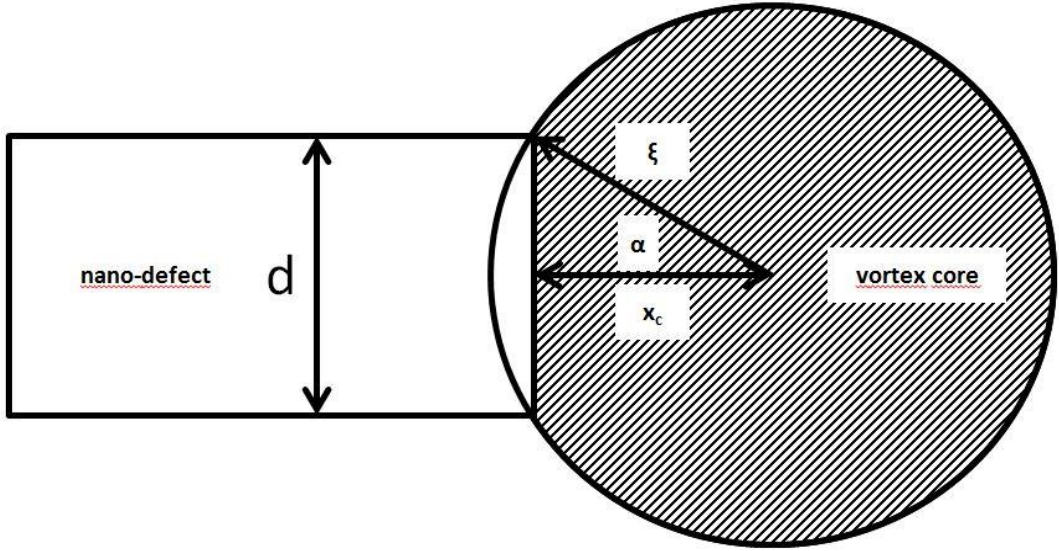

Fig. 4. Deflection of the pancake vortex core against the capturing center onto the critical length $x_c$.

In the initial state, it is for $x=0$ relation 8 reduces to Eq. 3. Physical meaning of the critical distance $x_C$ separating both regions is presented in Fig. 4. For the deflection of the pancake vortex core on the distance $x>x_c$, system energy of the vortex core captured on the nano-sized defect increases to the value:

$$U_2(x) = \frac{\mu_0 H_c^2}{2} l \left(-\frac{\alpha}{\pi}\pi\xi^2 + x\xi \sin\alpha\right) = \frac{\mu_0 H_c^2}{2} l \left(x\xi\sqrt{1-(\frac{x}{\xi})^2} - \xi^2 \arcsin\sqrt{1-(\frac{x}{\xi})^2}\right) \quad (9)$$

For $x = x_c$ relations 8 and 9 are convergent and reduce to the form:

$$U_1(x_c) = \frac{\mu_0 H_c^2}{2} l\xi^2 \left(\frac{d}{2\xi}\sqrt{1-(\frac{d}{2\xi})^2} - \arcsin\frac{d}{2\xi}\right) \quad (10)$$

Derived in geometrical approach expressions for the pinning potential allow then to calculate the pinning forces of interaction in first, it is for $x<x_c$ and second $x>x_c$ regions:

$$F_{p1} = -\frac{\partial U_1(x)}{\partial x} = -\frac{\mu_0 H_c^2}{2} ld \quad (11)$$

$$F_{p2} = -\frac{\partial U_2(x)}{\partial x} = -\mu_0 H_c^2 l\xi \sqrt{1-(\frac{x}{\xi})^2} \quad (12)$$

For pancake vortex shift on the distance $x = x_c$ both forces are equal. Basing on these considerations an energy barrier for pinning forces is calculated. For the deflection of the pancake vortex against the capturing center in the first region, it is for $x<x_c$ the pinning potential barrier reaches the value:

$$\Delta U_1(x) = \frac{\mu_0 H_c^2}{2} ldx \quad (13)$$

while for larger shift it is given by:

$$\Delta U_2(x) = \frac{\mu_0 H_c^2}{2} l\xi^2 \left(\arcsin(\frac{x}{\xi}) - \frac{\pi}{2} + arcsin(\frac{d}{2\xi}) + \frac{x}{\xi}\sqrt{1-(\frac{x}{\xi})^2} + \frac{d}{2\xi}\sqrt{1-(\frac{d}{2\xi})^2}\right) \quad (14)$$

In limiting case of x=$x_C$ Eqs. 13 and 14 stitch together:

$$\Delta U_2(x_c) = \frac{\mu_0 H_c^2}{2} l\xi^2 \left(\arcsin(\frac{x_c}{\xi}) - \frac{\pi}{2} + arcsin(\frac{d}{2\xi}) + \frac{x_c}{\xi}\sqrt{1-(\frac{x_c}{\xi})^2} + \frac{d}{2\xi}\sqrt{1-(\frac{d}{2\xi})^2}\right) = \quad (15)$$

$$\frac{\mu_0 H_c^2}{2} l\xi^2 \left(\arcsin\sqrt{1-(\frac{d}{2\xi})^2} - \frac{\pi}{2} + arcsin(\frac{d}{2\xi}) + \frac{d}{\xi}\sqrt{1-(\frac{d}{2\xi})^2}\right) = \frac{\mu_0 H_c^2}{2} l\xi\, d\sqrt{1-(\frac{d}{2\xi})^2} = \Delta U_1(x_c)$$

where it has been used trigonometrical identity:

$$\arcsin\sqrt{1-(\frac{d}{2\xi})^2} = \frac{\pi}{2} - arcsin(\frac{d}{2\xi}) \quad (16)$$

Movement of the pancake vortex against the capturing center is caused by the Lorentz force, which breaks off captured vortices in magnetic field and during transport current flow. Influence of Lorentz force is included by additional potential energy:

$$U_L = -jBV_i x = -jB\pi\xi^2 l x \quad (17)$$

which leads then to the tilting of the potential barrier shape in the following manner:

$$\Delta U_{1L}(x) = \frac{\mu_0 H_c^2}{2} l d x - jB\pi\xi^2 x l \quad (18)$$

$$\Delta U_{2L}(x) = \frac{\mu_0 H_c^2}{2} l\xi^2 \left(\arcsin\frac{x}{\xi} - \frac{\pi}{2} + arcsin(\frac{d}{2\xi}) + \frac{x}{\xi}\sqrt{1-(\frac{x}{\xi})^2} + \frac{d}{2\xi}\sqrt{1-(\frac{d}{2\xi})^2}\right) - jB\pi\xi^2 x l$$

$$(19)$$

The derivative in the respect to *x* of the energy barrier *ΔU* allows to determine position of its maximum:

$$\frac{\partial \Delta U_{1L}(x)}{\partial x} = \frac{\mu_0 H_c^2}{2} l d - jB\pi\xi^2 l \quad (20)$$

$$\frac{\partial \Delta U_{2L}(x)}{\partial x} = \mu_0 H_c^2 l\xi\sqrt{1-(\frac{x}{\xi})^2} - jB\pi\xi^2 l \quad (21)$$

According to the relation:

$$\frac{\partial \Delta U_{2L}(x)}{\partial x}\Big|x_m = 0 \quad (22)$$

it is received then the result:

$$\mu_0 H_c^2 l\xi\sqrt{1-(\frac{x_m}{\xi})^2} - jB\pi\xi^2 l = 0 \quad (23)$$

We introduce new parameter $j_{c1}$, which fulfills the function of the critical current density connected with individual capturing process:

$$j_{c1} = \frac{\mu_0 H_c^2}{B\pi\xi} \tag{24}$$

For regularly arranged ordered defects, the current density flowing through the sample is dependent on the defects concentration, given by the distance between capturing centers $a$ and defects geometry. In calculations of the current-voltage characteristics this geometrical factor is simply described by the defect cross-section S. Increase of it describes enhancement of the interaction probability with vortices and from other side shortage of superconducting material. Average critical current density is in the following form taking into account already the defects concentration and their shape:

$$j_c = \frac{\mu_0 H_c^2}{B\pi\xi} \cdot \frac{S}{a^2} \cdot \left(1 - \frac{S}{a^2}\right) \tag{25}$$

Relation 23 reduces then to the equation determining the position of the maximum of the potential barrier in the function of normalized current density.

$$\sqrt{1 - \left(\frac{x_m}{\xi}\right)^2} = \frac{j}{j_c} \tag{26}$$

Eq. 26 we write in the form showing directly the influence on the position of maximum of energy barrier $x_m$ of the reduced current density $i = j/j_c$.

$$x_m = \xi\sqrt{1 - i^2} \tag{27}$$

Then equation describing the maximum of the potential barrier is:

$$\Delta U_2(x_m) = \frac{\mu_0 H_c^2}{2} l\xi^2 \left(arcsin\frac{x_m}{\xi} - \frac{\pi}{2} + arcsin\left(\frac{d}{2\xi}\right) + \frac{x_m}{\xi}\sqrt{1 - \left(\frac{x_m}{\xi}\right)^2} + \frac{d}{2\xi}\sqrt{1 - \left(\frac{d}{2\xi}\right)^2}\right) - jB\pi\xi^2 l x_m \tag{28}$$

which after applying the trigonometrical identity 16 transforms into new form dependent already on the reduced current density $i$:

$$\Delta U(i) = \frac{\mu_0 H_c^2}{2} l\xi^2 \left(-arcsin(i) + arcsin\left(\frac{d}{2\xi}\right) + \frac{d}{2\xi}\sqrt{1 - \left(\frac{d}{2\xi}\right)^2} - i\sqrt{1 - i^2}\right) \tag{29}$$

In especial case of large defects size, starting from the condition $d = 2\xi$ relation 29 reduces to the form:

$$\Delta U(i) = \frac{\mu_0 H_c^2}{2} l\xi^2 \left(-arcsin(i) + \frac{\pi}{2} - i\sqrt{1 - i^2}\right) \tag{30}$$

which fills macroscopically required condition of positive potential barrier without current

$$\Delta U(i = 0) = \frac{\pi\mu_0 H_c^2}{4} l\xi^2 > 0 \tag{31}$$

and vanishing potential barrier for critical current flow as it predicts following relation:

$$\Delta U(i = 1) = 0 \qquad (32)$$

For slightly disordered nano-defects local critical current density can deviate from mean value given by Eq. 25, so averaged reduced current changes between $i$-$\Delta i$ and $i$+$\Delta i$, which leads to the averaged potential barrier:

$$\Delta U(i)_{av} = \frac{\mu_0 H_c^2}{4\Delta i} l\xi^2 \left( -\int_{i-\Delta i}^{i+\Delta i} (\arcsin(i) + i\sqrt{1-i^2})\, di + \arcsin\left(\frac{d}{2\xi}\right) + \frac{d}{2\xi}\sqrt{1 - \left(\frac{d}{2\xi}\right)^2} \right) =$$

$$\frac{\mu_0 H_c^2}{4\Delta i} l\xi^2 \left( 2\Delta i \left( \arcsin\left(\frac{d}{2\xi}\right) + \frac{d}{2\xi}\sqrt{1-\left(\frac{d}{2\xi}\right)^2} \right) - \begin{bmatrix} i \cdot \arcsin(i) + \sqrt{1-i^2} \\ -0.33(1-i^2)^{1.5} \end{bmatrix}_{i-\Delta i}^{i+\Delta i} \right) \qquad (33)$$

Additionally the elasticity energy of the vortex lattice is included into these considerations, which deformed by the flux capturing, is enhanced proportionally to the square of the shift of vortex from its equilibrium position in the lattice. Such behavior has been described by the energy potential given by simplified formula:

$$U_{el} = \frac{C\varepsilon^2}{2} V_i = \alpha_e\, (\xi - x)^2 \qquad (34)$$

where $V_i = \pi\xi^2 l$ is volume at which the deformation occurs, $C$ is module of elasticity, introduced in Eq. 1, while $\varepsilon = \xi - x$. Potential barrier calculated as the difference in elasticity energy of lattice between pancake vortex in initial position and deflected from vortices lattice is:

$$\Delta U_{el} = \alpha_e\, x(x - 2\xi) \qquad (35)$$

For low elasticity constant it is $\alpha_e \ll \mu_0 H_c^2 l$ relation 29 is retained and leads to the new expression for the total pinning potential barrier $\Delta U(i)_t$:

$$\Delta U(i)_t = \frac{\mu_0 H_c^2}{2} l\xi^2 \left( \begin{array}{c} -\arcsin(i) + \arcsin\left(\frac{d}{2\xi}\right) \\ +\frac{d}{2\xi}\sqrt{1 - \left(\frac{d}{2\xi}\right)^2} - i\sqrt{1-i^2} \end{array} \right) + \qquad (36)$$

$$\alpha_e\, \xi^2(\sqrt{1-i^2} - 2)\sqrt{1-i^2}$$

For larger value of the elasticity constant it appears the relation joining the position of the maximum of energy barrier with the reduced current density:

$$\frac{x_m}{\xi} = \frac{2\,\alpha_e i}{\mu_0 H_c^2 l} + \sqrt{1 - i^2 - \frac{4\,\alpha_e\, i}{\mu_0 H_c^2 l}} \qquad (37)$$

Above considerations deal to the case of the vortices captured initially on the depth equal to the coherence length $\xi$. Such configuration is preferred especially in superconductors with massive pinning centers because it should allow among other still for the flow of the superconducting shielding currents, around the vortex core, due to the proximity effect and keep this way the vortex structure.

3. Influence of initial pancake vortex position on the potential barrier form

Now we discuss second case of vortices captured totally onto arbitrary depth inside the nano-defect longer than *2ξ*, as it shows Fig. 5. Such configuration can be favorable for ultrathin defects, when independently on the vortex position inside the normal defect superconducting effects are still hold up and shielding currents further will flow. Superconductivity will be then continuously induced in nano-defect by the proximity effect.

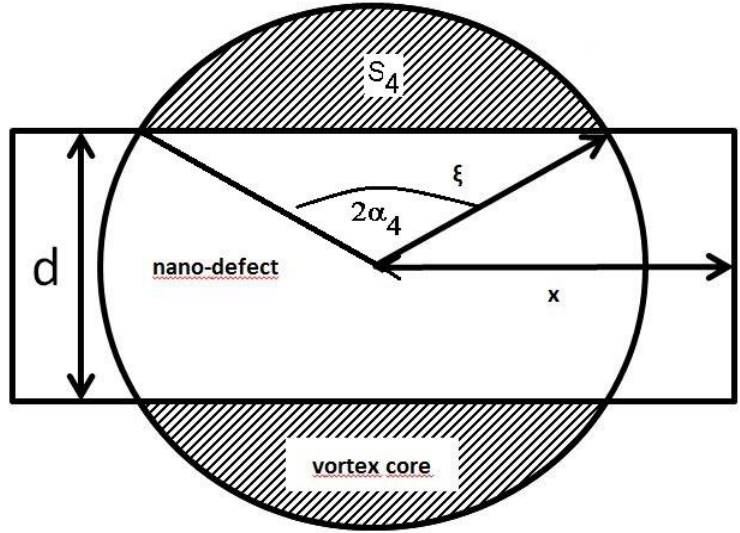

Fig. 5. Scheme of fully captured pancake vortex core inside thin nano-defect.

In this case energy of initial state connected with the capturing of the pancake type vortex is given as the function of the following dependences, in which used symbols are explained in Fig 5.

$$\sin\alpha_4 = \frac{\sqrt{\xi^2 - \frac{d^2}{4}}}{\xi} \qquad \alpha_4 = \arcsin\frac{\sqrt{\xi^2 - \frac{d^2}{4}}}{\xi} \tag{38}$$

$$S_4 = \xi^2 \arcsin\frac{\sqrt{\xi^2 - \frac{d^2}{4}}}{\xi} - \frac{d}{2}\sqrt{\xi^2 - \frac{d^2}{4}} = \frac{\pi\xi^2}{2} - \xi^2 \arcsin\frac{d}{2\xi} - \frac{d}{2}\sqrt{\xi^2 - \frac{d^2}{4}}$$

Then energy of the initial state for fully captured pancake vortex, independently on its real position, filling the condition *x < -ξ* is:

$$U(-\xi) = \frac{-\mu_0 H_c^2 l}{2}\left[2\xi^2 \arcsin\frac{d}{2\xi} + d\sqrt{\xi^2 - \frac{d^2}{4}}\right] \tag{39}$$

For intermediate shift of the pancake vortex core outside of the nano-sized defect described by the following range of the vortex deflection:

$$-\xi \leq x \leq -\xi\sqrt{1-\left(\frac{d}{2\xi}\right)^2} = x_{c1} \quad (40)$$

and shown in Fig. 6 enhancement of the energy of system of captured vortex is described by the relation 42, in the derivation of which has been used Eq. 41:

$$S_5 = \pi\,\xi^2\frac{\beta}{\pi} + x\sqrt{\xi^2 - x^2} = \xi^2 arcsin\frac{\sqrt{\xi^2-x^2}}{\xi} + x\sqrt{\xi^2 - x^2} \quad (41)$$

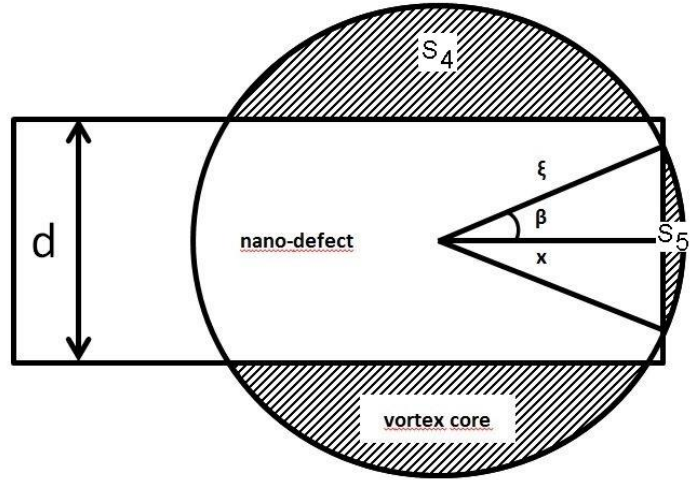

Fig. 6. Intermediate position of the captured pancake vortex.

and dependence $sin\beta = \sqrt{1-\left(\frac{x}{\xi}\right)^2}$ , while $S_4$ is given by Eq. 38. Finally pinning potential energy in this intermediate range of the vortex movement is expressed by the relation:

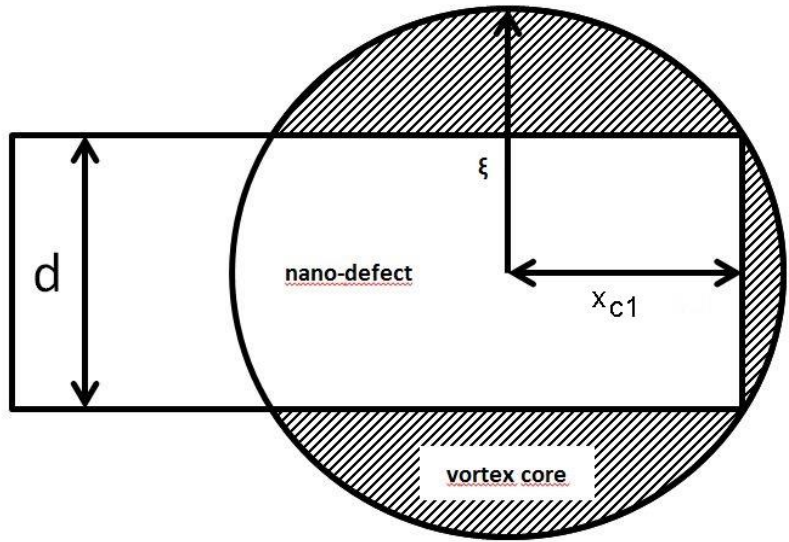

Fig. 7. Limiting shift of pancake vortex core leading to potential energy still in the form of Eq. 42.

$$U_1(x) = \frac{\mu_0 H_c^2 l}{2} \begin{bmatrix} -2\xi^2 \arcsin\dfrac{d}{2\xi} + \\ \xi^2 \arcsin\dfrac{\sqrt{\xi^2 - x^2}}{\xi} - d\sqrt{\xi^2 - \dfrac{d^2}{4}} \\ + x\sqrt{\xi^2 - x^2} \end{bmatrix} \qquad (42)$$

Graphical interpretation of limiting shift $x_{c1}$ of the vortex filling still this condition is shown in Fig. 7. For larger vortex movement described by the relation:

$$x_{c1} = -\xi\sqrt{1 - \left(\frac{d}{2\xi}\right)^2} \leq x \leq \xi\sqrt{1 - \left(\frac{d}{2\xi}\right)^2} \qquad (43)$$

areas of corresponding regions shown in Fig. 8 are: $S_6 = d\left(\xi\sqrt{1 - (\frac{d}{2\xi})^2} + x\right)$, while $S_4$ and $S_5$ were determined previously.

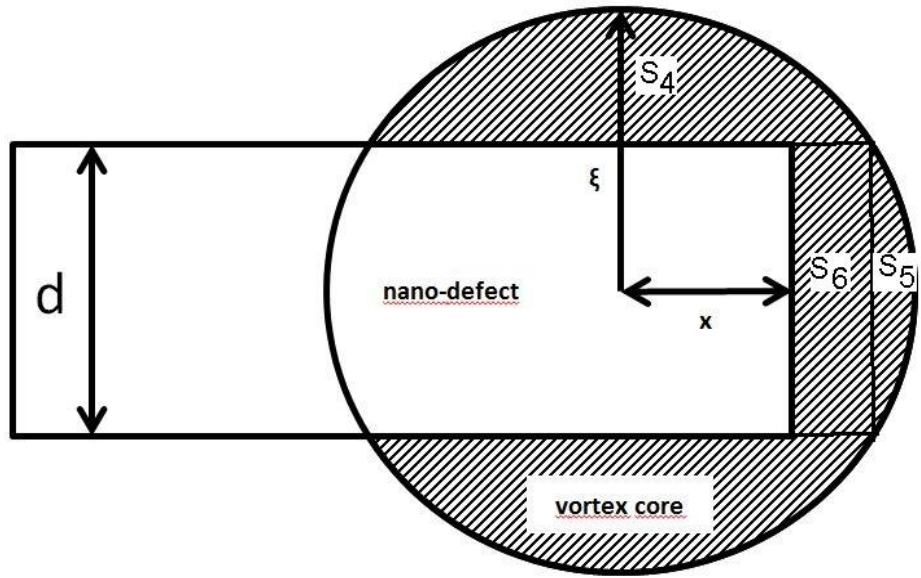

Fig. 8. Subsequent configuration of movement of captured pancake vortex.

Pinning potential energy of shifted vortex is given then by the relation 8, while for larger $x$ potential energy is given by Eq. 9. Expressions for the potential barrier height, which determine the current-voltage characteristics have been derived next and are given below for each of the considered cases, taking into account additionally Lorentz forces potential:

$$\Delta U_1(x) = \frac{\mu_0 H_c^2 l}{2}\left[\xi^2 \arcsin\frac{\sqrt{\xi^2-x^2}}{\xi} + x\sqrt{\xi^2-x^2}\right] - jB\pi\xi^2 l\,(x+\xi) \qquad (44)$$

$$\Delta U_2(x) = \frac{\mu_0 H_c^2 l}{2}\left[\xi^2 \arcsin\frac{d}{2\xi} + \frac{d\xi}{2}\sqrt{1-(\frac{d}{2\xi})^2} + dx\right] - jB\pi\xi^2 l\,(x+\xi) \qquad (45)$$

$$\Delta U_3(x) = \frac{\mu_0 H_c^2 l\,\xi^2}{2}\left[\begin{array}{l}\arcsin\frac{x}{\xi} + \frac{x}{\xi}\sqrt{1-\left(\frac{x}{\xi}\right)^2} + 2\arcsin\frac{d}{2\xi} \\ +\frac{d}{\xi}\sqrt{1-\left(\frac{d}{2\xi}\right)^2} - \frac{\pi}{2}\end{array}\right] - jB\pi\xi^2 l\,(x+\xi) \qquad (46)$$

Then analogously to the method described previously we can transform the relation describing the position of maximum of energy barrier $\Delta U(x_m)$ into transport current representation using the relation $x_m = \xi\sqrt{1-i^2}$ :

$$\Delta U_3(i) = \frac{\mu_0 H_c^2 l\,\xi^2}{2}\left[\begin{array}{l}-\arcsin i - i(\sqrt{1-i^2}+2) + 2\arcsin\frac{d}{2\xi} \\ +\frac{d}{\xi}\sqrt{1-\left(\frac{d}{2\xi}\right)^2}\end{array}\right] \qquad (47)$$

Current-voltage characteristics has been calculated in this approach after inserting potential barrier height $\Delta U$ into the flux creep equation:

$$E = B\omega\,a\left(e^{\frac{-\Delta U(0)(1+i)}{k_B T}} - e^{\frac{-\Delta U(i)}{k_B T}}\right) \qquad (48)$$

which describes flux creep forward and backward processes, with hoping frequency ω, $k_B$ is Boltzmann's constant. Results of calculations are shown in Fig. 9 and indicate decrease of critical current with applied magnetic field, as it is experimentally observed in Fig. 1.

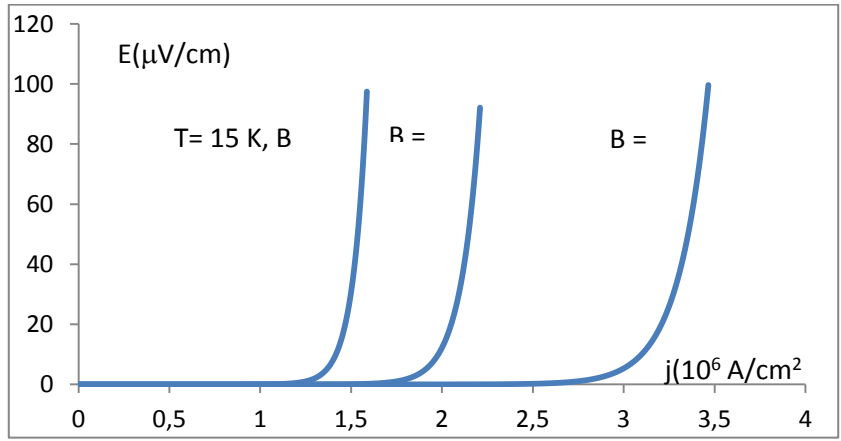

Fig. 9. Theoretically calculated current-voltage characteristics of HTc superconductor in static magnetic field.

Materials parameters used in calculations shown in Fig. 9 are: $T_C = 108$ K, $\xi(0) = 2{,}57$ nm, $\lambda(0) = 200$ nm, $B_{irr} = 22$ T, width of the pinning center d = 2 nm, length 7 nm, while thickness of defect created by fast neutrons irradiation process is equal to thickness of pancake vortex in superconducting layer l = 10 nm. Such sizes clearly reflect the thin pancake vortex structure preferring the individual capturing and allowing to neglect the flux cutting and bending processes as well as flux bundle capturing. Temperature and magnetic induction values are given in Fig. 9. According to above dimensions the configuration of initial state shown in Fig. 5 and Eq. 47 were used in calculation process.

Third considered case shown in Fig. 10 is characterized by value of parameter $x_0 = -\xi/2$. Then expression on potential barrier in the current representation brings the form:

$$\Delta U_3(i) = \frac{\mu_0 H_c^2 l\, \xi^2}{2} \left[ \begin{array}{l} -\arcsin(i) - i(\sqrt{1-i^2} + 1) \\ +\arcsin\dfrac{d}{2\xi} + \dfrac{d}{2\xi}(1 + \sqrt{1 - \left(\dfrac{d}{2\xi}\right)^2}) \end{array} \right] \quad (49)$$

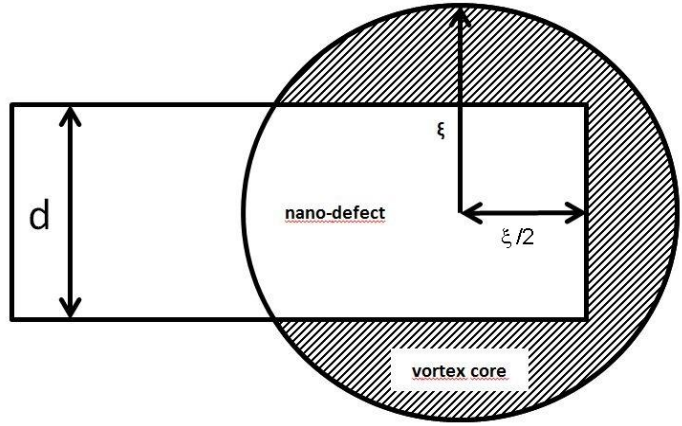

Fig. 10. View of captured pancake vortex on intermediate distance -$\xi$/2.

We should notice however that potential barrier in the present model does not fill the condition of vanishing for critical current flow. The renormalizing procedure has been therefore applied, similar to used in the collective flux creep theory [9], allowing to solve this problem. For initial pancake vortex positions $x_0 = 0$, -$\xi$/2 and -$\xi$ the expressions for potential barrier versus transport current density are equal then respectively:

$$\Delta U(i)_t = \frac{\mu_0 H_c^2}{2} l\xi^2 \begin{pmatrix} -\arcsin(i) + \arcsin\left(\frac{d}{2\xi}\right) \\ + \frac{d}{2\xi}\sqrt{1-\left(\frac{d}{2\xi}\right)^2} \end{pmatrix} - i\left(\sqrt{1-i^2} + \arcsin\left(\frac{d}{2\xi}\right) + \frac{d}{2\xi}\sqrt{1-\left(\frac{d}{2\xi}\right)^2} - \frac{\pi}{2}\right) +$$
$$\propto_e \xi^2(\sqrt{1-i^2} - 2)\sqrt{1-i^2} \tag{50}$$

$$\Delta U(i) = \frac{\mu_0 H_c^2 l\, \xi^2}{2}\left[\begin{array}{l} -\arcsin(i) + \arcsin\dfrac{d}{2\xi} \\ + \dfrac{d}{2\xi}\left(1+\sqrt{1-\left(\dfrac{d}{2\xi}\right)^2}\right) + i\left(\dfrac{\pi}{2} - \dfrac{d}{2\xi}\left(1+\sqrt{1-\left(\dfrac{d}{2\xi}\right)^2}\right) - \arcsin\left(\dfrac{d}{2\xi}\right) - \sqrt{1-i^2}\right) \end{array}\right] \tag{51}$$

$$\Delta U(i) = \frac{\mu_0 H_c^2}{2} l\xi^2 \begin{pmatrix} -\arcsin(i) + 2\arcsin\left(\frac{d}{2\xi}\right) + \frac{d}{\xi}\sqrt{1-\left(\frac{d}{2\xi}\right)^2} \\ +i\left(\frac{\pi}{2} - 2\arcsin\left(\frac{d}{2\xi}\right) - \frac{d}{\xi}\sqrt{1-\left(\frac{d}{2\xi}\right)^2} - \sqrt{1-i^2}\right) \end{pmatrix} \tag{52}$$

The results of calculations of the influence of reduced transport current on potential barrier in reduced unit are given in Figs. 11-12. Fig. 11 presents the potential barrier in reduced units in the function of current for various dimensions of pinning centers, which correspond to various sizes and energies of irradiating heavy ions in superconducting accelerators [11], while Fig. 12 concerns various initial states of pancake vortex.

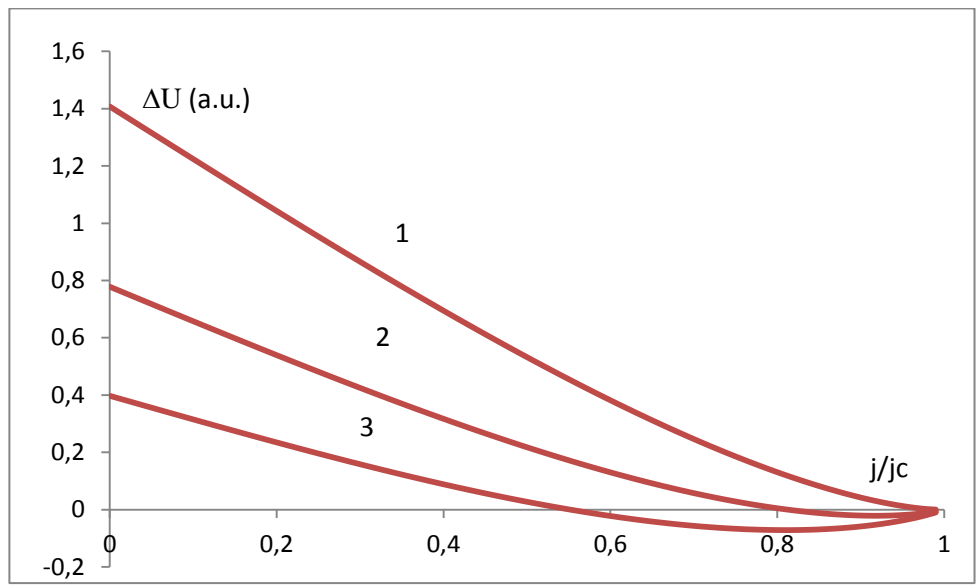

Fig. 11. Influence of reduced current *i* on potential barrier *ΔU* for half-pinned initially vortex versus size of nano-defect: (1) $d/2\xi = 0{,}8$, (2) $0{,}4$, (3) $0{,}2$

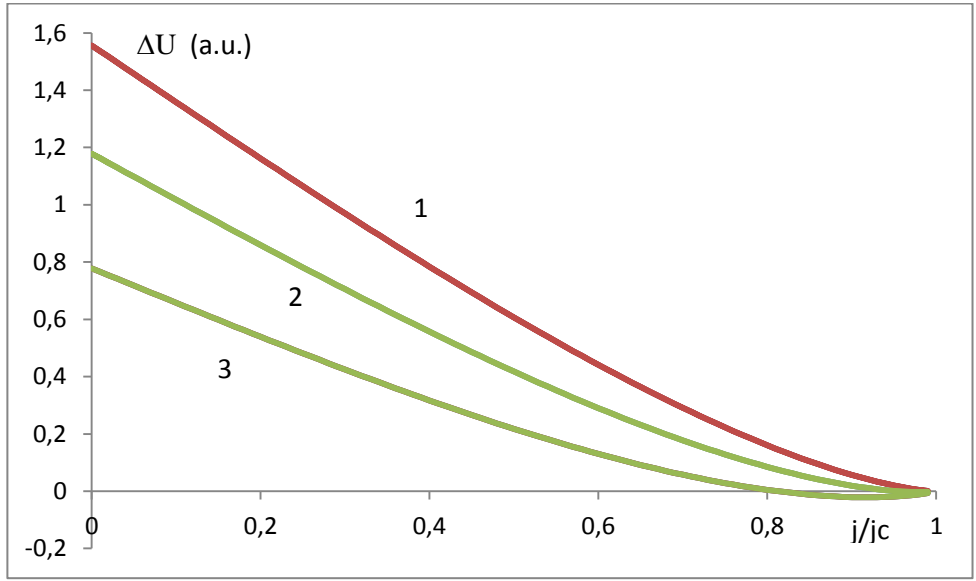

Fig. 12. Influence of initial position of captured vortex on reduced current versus potential barrier $\Delta U$ dependence for $d/2\xi = 0,4$, (1) $x_0 = -\xi$, (2) $x_0 = -\xi/2$, (3) $x_0 = 0$.

Finally Fig. 13 presents the result of calculations according to presented model of influence of concentration of nano-defects created by fast neutrons irradiation on critical current density of HTc superconductor in full irradiation concentration range starting from pure superconductor up to fully irradiated sample, in the function of magnetic induction. It was assumed here that irradiation penetrating superconducting layers creates regularly ordered nano-defects of width 7 nm. Multilayered structure of HTc superconductors composed from superconducting and buffer layers has been taken into account at calculation effective cross-section used for determination of the current density. Superconducting materials parameters used in calculations as critical temperature, coherence length, penetration depth, irreversibility field were the same as previously used for results shown in Fig. 9. The increase of the critical current in initial irradiation range is in qualitative agreement with experiments [11] performed on $Nb_3Sn$ wires using source of 65 MeV and 24 GeV protons.

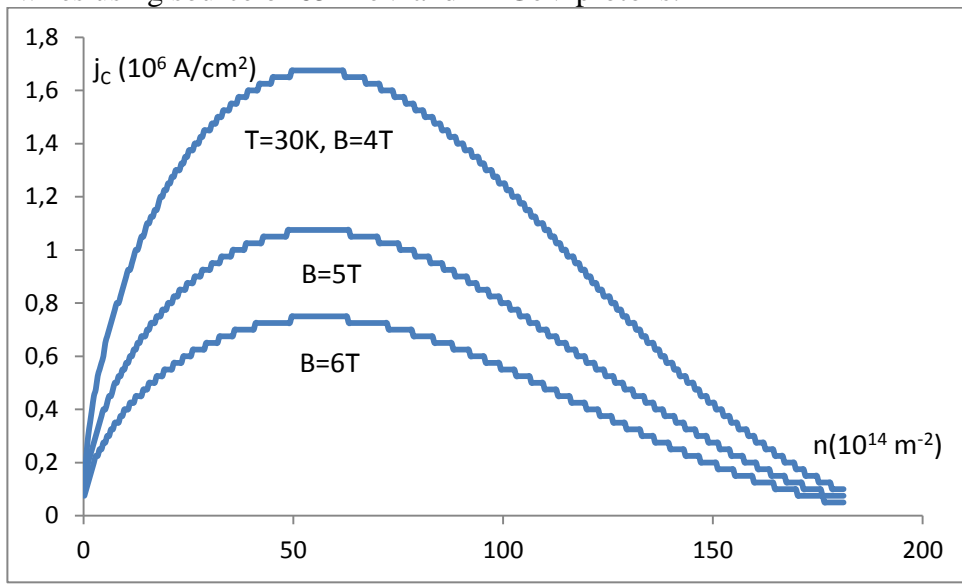

Fig. 13. Calculated critical current density dependence on fast neutrons irradiation surface concentration (*n*) in the function of magnetic induction.

Energy of irradiated particles can influence the size and depth of nano-defects. The maximal critical current for the irradiation dose of the range 25% of maximal dose is in agreement with numerical calculations presented in [12]. Also current-voltage characteristics presented in Fig. 9 are in qualitative accordance with data given in [12].

Conclusions

In conclusions it can be stated that received in this model expressions on the pinning potential barrier allowed to calculate the shapes of I-V curves of multilayered superconductors and dependence of critical current on nano-defects concentration, which are in qualitative agreement with available experimental data [11] and results of advanced numerical calculations given in [12]. More advanced geometries and physical cases will be considered in subsequent papers.